# VIZGR: COMBINING DATA ON A VISUAL LEVEL

Daniel Hienert, Benjamin Zapilko, Philipp Schaer, Brigitte Mathiak
*GESIS – Leibniz Insitute for the Social Sciences, Lennéstr. 30, Bonn, Germany*
*daniel.hienert@gesis.org, benjamin.zapilko@gesis.org, philipp.schaer@gesis.org, brigitte.mathiak@gesis.org*

Abstract: In this paper we present a novel method to connect data on the visualization level. In general, visualizations are a dead end, when it comes to reusability. Yet, users prefer to work with visualizations as evidenced by WYSIWYG editors. To enable users to work with their data in a way that is intuitive to them, we have created Vizgr. Vizgr.com offers basic visualization methods, like graphs, tag clouds, maps and time lines. But unlike normal data visualizations, these can be re-used, connected to each other and to web sites. We offer a simple opportunity to combine diverse data structures, such as geo-locations and networks, with each other by a mouse click. In an evaluation, we found that over 85 % of the participants were able to use and understand this technology without any training or explicit instructions.

## 1 INTRODUCTION

Presenting data in an appealing or at least understandable way is a common task in the modern working environment. While the classical business application is presenting company data to support decision making, it has also become important in any teaching job or marketing. The reason for this is that raw data, such as can be found in databases are collected very easily, but the pure tables are hard to understand.

Since it is such a widespread task, it is clear that most of the people that are making such visualizations are not experts in data visualization or statistics. They are experts on the data they want to present and this data is typically complex and interconnected.

The classical spreadsheet approach often does not adequately represent neither the complexity nor the interconnectedness. Also, the people who are typically using such applications are not willing to learn complex data-centered techniques to represent this interconnectedness, as they are typically just modifying the data so it can be visualized more appropriately.

The solution to this dilemma is the option to interconnect the visualizations directly. Since the visualized data is what users are thinking of anyway when handling the large arrays of data, it is only logical to allow them to work on the visualizations directly.

In Vizgr, we allow users to interconnect their data visualizations like they could any other object on the Internet. The web-based framework allows sharing of visualizations (connecting to people), connecting visualizations with each other on a data level and connecting visualizations with web sites to fully embed it into the World Wide Web.

As our survey shows, the user acceptance and usability of Vizgr is very high. Most of the participants can imagine scenarios in which they would like to use this tool in their daily work.

In the next few sections, we will, first, discuss related products and ideas. In section three, we will give an overview on the capabilities and technical details of Vizgr. We will proceed in section four with some use cases we have been studying. In section five we present the survey and the results of the survey we have conducted, to investigate both the usability of the tool and the usability of the idea. We conclude with some final remarks on future work, we have planned.

## 2 RELATED WORK

Vizgr integrates the key ideas of collaborative sharing of visualizations on the web, the coordinated views of multiple visualizations on one web page and extends the concept to manual, semiautomatic and automatic linking of visualizations for browsing and coordination purposes. In this section, we first present related systems and their key ideas and conclude how Vizgr distinguishes to them.

## 2.1 Visualization on the Web

IBM Many Eyes (Viegas et al., 2007) and Swivel are online tools for sharing data and visualizations on the web. The user can upload data, choose a visualization type and create a visualization that can be viewed and commented by the community. By integrating an HTML snippet the visualization can be embedded on other sites or blogs. The underlying data set can be reused by other users to build their own visualizations. (Heer et al., 2009) gives an overview of these and other online visualization tools, their functionality and impacts.

VisGets (Dörk et al., 2008) uses different visualizations to show and filter retrieved web resources in several dimensions like time, location and topic. Based on the concept of dynamic queries, results can interactively be filtered by manipulating the visualizations. VisGets also implements coordinated interactivity. Hovering with the mouse over a visual element highlights all related elements in the visualizations and in the result list. The new introduced approach of *Weighted Brushing* is used to highlight strongly related items more than weakly related ones.

Dashiki (McKeon, 2009) is a wiki-based collaborative platform for creating visualization dashboards. Users can integrate visualizations that contain live connections to data sources. Data sets are embedded into data pages by a special markup, via Copy&Paste from spreadsheets or by an URL. Live data is dynamically fetched and stripped from formatting tags, so the user can wrap the content with needed markup. Dashiki uses a simple technical approach for coordinated selecting among multiple views. Simple attribute-value pairs are propagated to all visualizations via JSON format.

Exhibit (Huynh et al., 2007) is a lightweight framework for easy publishing of structured data on the web. Users can import data via JSON, which is presented on the web page in different views including maps, table, thumbnails and timelines.

## 2.2 Coordinated Views

The simultaneous display of the same data structure in several different views was first defined by (Baldonado et al., 2000). They set up a model for coordinated multiple views and provide guidelines for not disrupting the positive effect through increased complexity. The main idea is that data in different views can be linked. If data is selected in one view, it is also highlighted in other views (brushing-and-linking).

North & Shneiderman (North and Shneiderman, 2000) provide an alternative visualization model which is based on the relational data model. The system Snap (North et al., 2002) is an implementation of this model. It allows the user to select databases and assign visualizations. In a second step, the user can then connect different visualizations and generate coordinated visualizations. Highlighting or other actions are coordinated between the different views. For example, if data is selected in one view it is also selected in the other views.

VisLink (Collins and Carpendale, 2007) is a system for revealing relationships amongst different visualizations. Multiple visualizations are drawn on 2D planes and can be placed in a 3D space. Relationships are displayed between them by propagating edges from one visualization to another. Relationships, connections and patterns between these visualizations can be explored by several interaction techniques.

## 2.3 Unique features of Vizgr

Vizgr is similar in certain aspects to the presented systems, but differs in some others. On a basic level Vizgr can create visualizations like the aforementioned systems. However, most tools concentrate on one or some information types like tabular data, text, maps and so on; in contrast Vizgr supports heterogeneous information types like tabular data, text, locations, events and network data and heterogeneous visualizations like business graphics, tag clouds, maps, time lines and network graphs in one tool. Vizgr supports the user with easy creation of visualizations with different forms, possibilities for copy & paste from spreadsheets and automatic data import from Wikipedia.

Similar to Dashiki and Exhibit Vizgr supports the integration of several visualizations on one webpage. But in contrast to those systems, the selection of visualizations in Vizgr is based on the relationships between visualizations.

Vizgr has a similar approach to highlight related visual items in different visualizations to VisGets or Dashiki. But in Vizgr the coordinated view is not only based on the same data items in different visualizations, but also on relationship data.

A completely new aspect in Vizgr is the support for the manual, semiautomatic and automatic mapping of heterogeneous visualization types and from visualizations to websites. This allows the creation of networks of visualizations for browsing and coordination purposes.

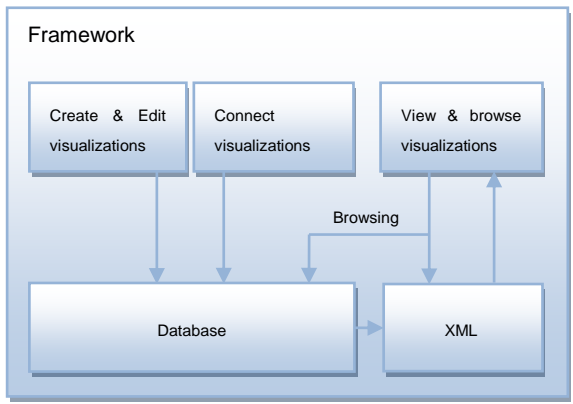

Figure 1: Vizgr system overview.

## 3  VIZGR

In the following section we describe the architecture of Vizgr and the workflows to create and map visualizations to each other and to websites. We introduce the browsing approach and the possibility of coordinated views.

Vizgr includes components to create, view, modify, save and connect visualizations. Figure 1 gives an overview of the system's architecture. All modules are integrated in a web framework, which is implemented in PHP. The creation and editing of visualizations can be done in one single HTML form, the workflow and exact details are described in section 3.1. The entered data is stored in the user session and in the Vizgr database. In a second step the user can either connect two visualizations to each other or a visualization to websites. The workflows for connecting visualizations are further described in section 3.2. Connection data is also stored in the user session and in the database. Visualization data and properties are committed to the visualization component via XML. The framework creates an XML file that contains all information that is necessary for the visualization component to create the visualization, set the linking buttons and mark individual items.

The core component for building and viewing the visualization is a Flash application that is implemented in ActionScript 3. We have chosen Flash, because it is widely distributed, does not need any preloading time to start a virtual machine and offers all available possibilities to implement advanced graphics and user interaction. The module parses the XML and builds the appropriate visualization.

### 3.1  Creating Visualizations

The user can create different visualizations with the help of an HTML form. For creating a visualization the user has to go through four steps: (i) enter a title, (ii) enter a description, (iii) enter data and (iv) choose a visualization type. As a result, a preview of the created visualization is shown. All steps are accomplished in one single HTML form. This makes it possible to edit and correct certain entries and to see the result immediately in the preview. Title and description are metadata fields the user can fill out to identify and describe the visualization. Data can be entered with (i) different data input templates belonging to the information type, (ii) by copy and paste from spreadsheets or (iii) by automatically loading the data from Wikipedia or the DBpedia database.

For small amounts of data the user can enter the information manually. The framework offers data input templates for (i) tabular data, (ii) text, (iii) locations, (iv) events and (v) network data. The tabular data template is structured like an excel sheet. The user can enter different attributes and the respective data. Textual data can be entered simply by copying it into a text field. The location template offers fields for title, description and address details like street, house number, post code, city and country. The framework has a built in geocoder to add latitude and longitude information to the record. Events can be entered with the attributes title, description, start and end date. The network data template is a simplified table data template with three columns. Related nodes can be entered in columns one and two. Column three is an optional field for entering the relationship between nodes.

Copy and Paste from spreadsheets is appropriate for large data collections already available in CSV, like for example finance data from Yahoo. For locations, events and network data the user has to format the columns in a certain order. Then data can just be marked in the spreadsheet, copied to the clipboard and pasted in the form.

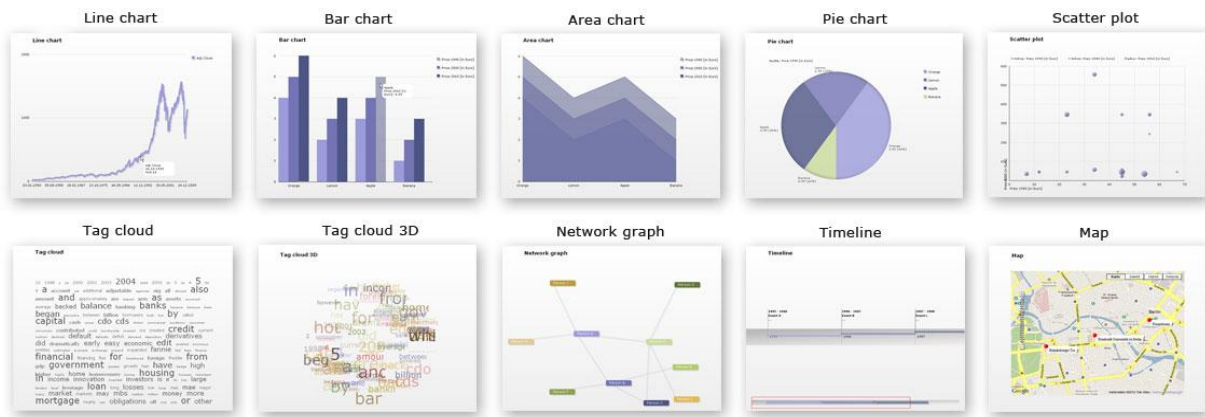

Figure 2: An overview of visualizations that can be created with Vizgr: for tabular data: line chart, bar chart, area chart, pie chart, scatter plot; for text: tag cloud and 3D tag cloud; for network data: a network graph; for events: a timeline; and for locations: a Google map.

For the information types text, location and event exist the possibility to load the data directly from Wikipedia and the DBpedia database that offers structured information from Wikipedia articles. To get the text of a Wikipedia article the user can enter a topic in the search field. Via an autocomplete list the user can identify existing topics and choose one. With a click on a button the text is loaded and can be used for tag clouds. The benefit is that links to other Wikipedia articles are automatically extracted and provided by linking buttons (further described in section 3.3). For locations on a map the user can enter different location names in the fields. Coordinates, description and links are loaded from DBpedia. The user is again supported by an autocomplete list of relevant topics. For events in a specific time period the user can enter a start and an end date. Vizgr checks for appropriate events in the DBpedia database that can be visualized in a timeline.

Depending on the chosen information type, the framework offers appropriate visualization types. Figure 2 shows the different types of visualization which can be created. For tabular data as follows: line chart, area chart, bar chart, pie chart, scatter plot; for text: tag cloud and 3D tag cloud; for locations: a Google map; for events: a timeline; and for network data: a network graph. Users can choose visualization types and see the created visualization in the preview section.

## 3.2 Connecting Visualizations

In order to connect two visualizations, some kind of relationship data is needed. This can be automatically derived, e.g. from relational data in a database, an RDF store or by a matching attribute/value pair in a table. Also an editor for entering and editing the relations by hand is provided. The editor is directly based on the created visualization and is thus seamlessly integrated into the workflow of creating and viewing visualizations. Vizgr supports the connection of all its visualization types.

### 3.2.1 Connecting Two Visualizations

Graphical objects of two different visualizations can be manually connected using the Mapping Editor. The system supports the user by making suggestions for possible connections.

Graphical objects of the different types of visualization are, for example, a location marker on a map or a bar in a bar chart. A graphical object or glyph represents several data fields or properties in one simple graphical representation. The Mapping Editor utilizes this to simplify the mapping process. The user does not have to deal with the complex information structure on a data level, but can select graphical objects directly in the visualization. For the simple identification of a graphical object, the user receives additional information in a popup window by hovering with the mouse.

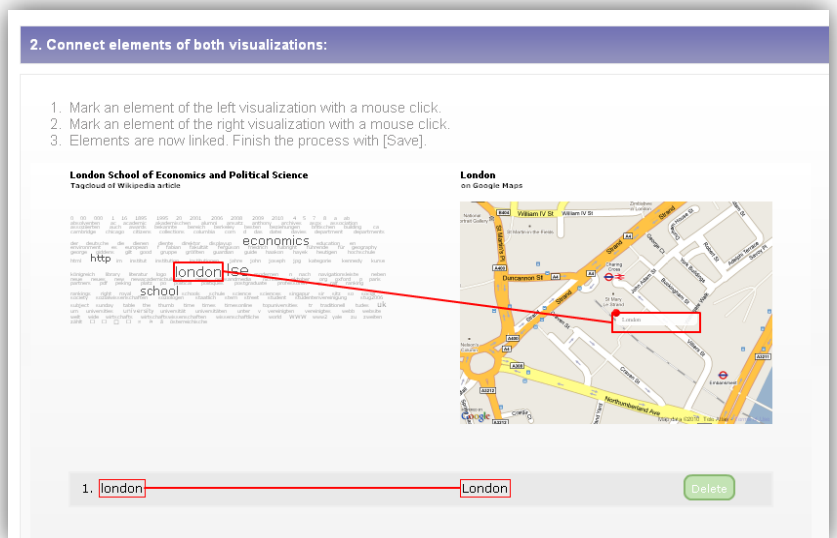

Figure 3: The Mapping Editor: mapping the word London in a tag cloud to the place on a map.

The user interface is organized as follows: at the top, the two chosen visualizations are shown side by side. The visualizations appear with the same functionality as in the view modus, this means that information for the graphical object is available by hovering with the mouse. A list of the connected graphical objects can be found below the visualizations. On the left side, each record displays the graphical object value of the origin visualization, and on the right side, the visual object value of the target visualization. A button is available to delete each record. There are buttons for *Save*, *Cancel* and *Suggest Mapping* actions at the bottom of the list. Figure 3 gives an overview of the Mapping Editor.

The first step in the mapping workflow is to choose the visualizations to be connected. To create a connection, a graphical object from the left or right visualization is selected. Objects that are available for selection are marked with a thin red border when moused over. Clicking the mouse selects the object and it is then marked with a thicker, red border. Moving the mouse towards the target visualization makes a red connection line appear that will visually connect the origin and target object. Clicking the mouse a second time will select the target object. Once a connection has been created, it appears in the list and is visible in the visualizations as two marked objects connected with a line.

All connections are similarly colour-coded in the visualization and in the list. This allows connections to be easily identified, for example, to find them in the visualization in order to delete them from the list. Clicking *Save* completes the workflow. The two visualizations are now connected, and the mapping process can be continued for other visualizations.

All created connections can again be loaded into the Mapping Editor and edited.

### 3.2.2 Semiautomatic Mapping

The system can support the manual mapping process by making suggestions for connections between information items. By clicking on the button *Suggest Mapping* the system analyses the underlying data set and shows possible links in the visualization and in the mapping table. The user can check these mappings and delete unwanted links.

The algorithm for automatic search of mappings works by pre-processing the underlying data for both visualizations. For graphics with tabular data we build an array of attribute/value pairs in a simple format. The mappings are created on a graphical level, but visual elements need a different number of attribute/value pairs to be visually created and identified for different visualizations. For tag clouds, we take the whole text, for Google maps the title and description and so on. Text elements are again split into single words. The algorithm then checks for every array element if there is an equal element in the array of the other visualization.

For example, if we connect a network graph of persons with a visualization of a timeline with publications, for every network node value the system searches for any occurrence of the name in the title or description. Meaning that all names are compared to all publications and immediately connected if a match is found.

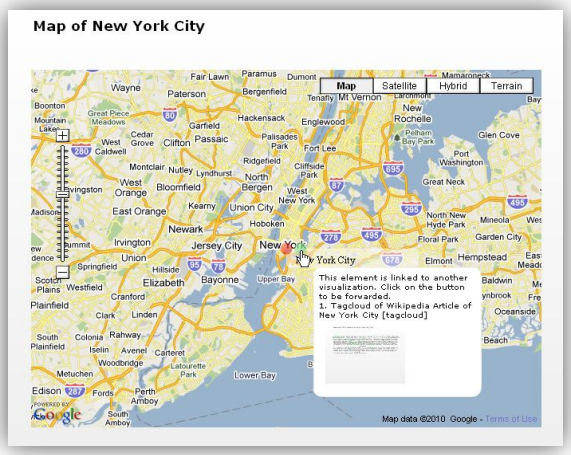

Figure 4: Hovering over the linking button of New York City shows a window with a link to the tag cloud of the Wikipedia article.

### 3.2.3 Connect a Visualization with Websites

Connecting items of visualizations to websites offers the possibility to create links from visual items to any resource on the web. This can be websites, URIs for literature references or visualizations in other portals. IBM ManyEyes e.g. supports addressing different states of visualizations via URL. The benefit is that visualizations and their visual items are included in the process of web hyperlinking and thus elevated from being fixed non-interactive illustrations.

Connecting visualizations to websites works similar to the workflow described for connecting two visualizations. The user chooses one visualization in the drop down list. Now he can mark any graphical object in the visualization with a mouse click. The object is then highlighted with a coloured border. For every marked object a list element appears where on the left side appears the object's name and on the right side one can enter a title and an URL for a website. The elements in the list are also visually connected with a frame in the same colour as in the visualization. The user can continue the process until he finishes it with a click on the save button. He then can have a look at the created connections in the visualization or start a new mapping process. The connections are listed in the personal area and can be edited later.

### 3.3 Browsing in Visualizations

All created visualizations appear in the gallery with title and thumbnails. Users choose a visualization which is displayed with functionality to edit, share and comment. If a graphical object in the source visualization is connected to another object in any target visualization or to a web resource, it is marked with a small green linking button (compare figure 4). Hovering over the button with the mouse pointer shows a window with the explanation that the element is linked to another visualization or website and with a click on the button one is forwarded. The window lists connected visualizations with title, visualization type and thumbnail and websites with title and URL. If the element has just one connection, clicking the button directly forwards the user to the connected visualization or website. Otherwise if a graphical object has connections to other, multiple visualizations or websites, a click on the button opens again a window listing the target visualizations by title and visualization type and websites by title and URL. The related visual element in the target visualization is highlighted with a red border. The target object is also marked with the linking button, which leads directly back to the origin visualization or other related visualizations and websites.

### 3.4 Coordinated Multiple Views

Beside the possibility to explore relations between visualizations by browsing from one to another in full view, the original and all related visualizations are also shown on the individual visualization page in half the scale. This way it is possible to see at once multiple related visualizations. Based on the approach of highlighting the same data item in different views, we choose an approach to highlight connected data in the different heterogeneous visualizations. Hovering the mouse pointer over a linked visual item in the original visualization highlights all connected items in the connected visualizations with a green frame (compare figure 5). If these items are linked to other items again, they send secondary events to all related visualizations that appear with a yellow frame. So, the user can interactively explore which visualizations are directly or indirectly connected to the chosen item. For example, when hovering with the mouse pointer over a person in a network graph, the connected location on the map is highlighted with a green frame. The location itself is connected with several events on a timeline that are highlighted with a yellow frame. The user can see that the person is indirectly connected with these events.

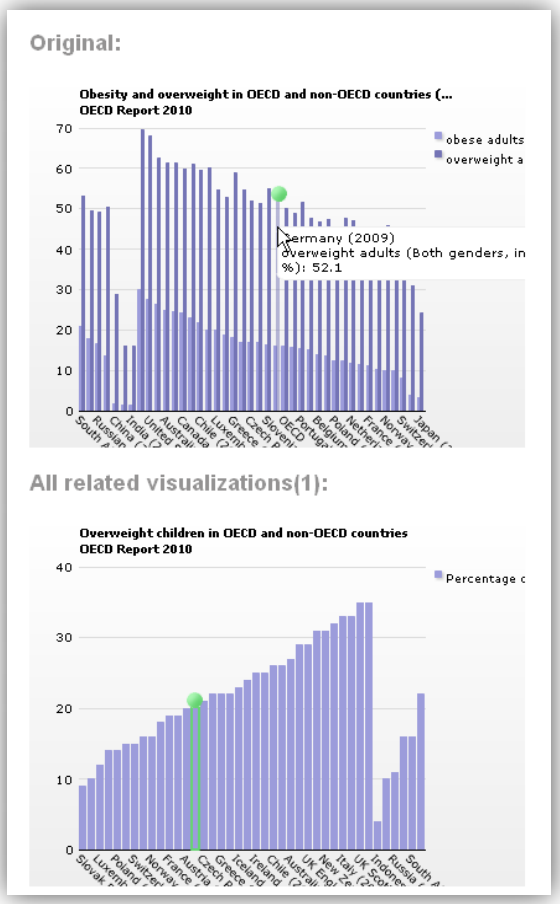

Figure 5: Coordinated Multiple View: Hovering with the mouse over a graphical object, highlights all related object in other views.

## 4 USE CASES

In this section we demonstrate the capabilities of Vizgr by giving two examples of visualization that can be produced within minutes. Simple mappings produce graphics showing relationships which may be retrievable in the web. But with Vizgr they are much easier to explore and produce. In our first example, we show relationships between 60 years stock prices and historical events on financial crisis. Relationships between minima and maxima of stock prices and related historical events can be explored with a mouse click. Our second example is also the basis scenario for our user study. We show the workflows to create a tag cloud and a map of chosen Wikipedia data and connect the visualizations to each other and to the web.

In the first example, we connect a line chart of historical data for S&P 500 stocks with historical events on financial crisis on a timeline. The stock data is taken from Yahoo, consists of about 15,000 rows of daily adjusted closing stock prices from 1950 to 2009. The data is entered via Copy & Paste from the provided CSV file. Selected historical events on financial crisis with title, details, start and end date is entered manually. In a second step, we connect the visualizations in the Mapping Editor. Both data sets have a date field, so the system propose a mapping automatically. The result is a line chart with historical stock prices connected with a timeline of historical events on financial crisis. The user can click on certain points of the line charts maxima and minima marked with the linking button. A click forwards to the appropriate historical event on the timeline which is highlighted with a red rectangle. A click on an event leads back to the appropriate stock price of that date. Figure 6 shows the line chart with historical stock prices and the timeline with historical events across financial crisis. The visualizations could be connected simple and fast and result in interactive linked visualizations that can be explored with a mouse click.

In our second example we show the workflow to create two visualizations from Wikipedia data, connect them to each other and to a website. First visualization is a tag cloud of the Wikipedia article on *London School of Economics*. We choose *Create Visualization* and enter a title and a description for the visualization. In the data input menu we choose *Wikipedia*. We now enter the first letters of *London School of Economics* and Vizgr proposes matching articles from Wikipedia in a drop down list. We select the entry with the mouse and click on *Load text from Wikipedia*. In the next step we choose *tag cloud* as a visualization type and click on *Create* to see the preview. The visualization has been created and can now be saved. Second visualization is a map with Wikipedia data from *London*. The workflow is similar: enter a title and a description, choose *Wikipedia*, then *Places*. We enter the first letters of *London*, choose the matching article from the drop down list and click on *Load places from Wikipedia*. *Map* is already chosen as a visualization type; we can click on create and save the visualization. Now both visualizations can be connected in the Mapping Editor. Therefore we choose *Connect Visualizations* from the main menu. We choose both visualizations from the drop down lists. In the Mapping Editor we click in the tag cloud on the word London and a second click on the location marker of London in the map. The visualizations now are already connected and the mapping can be saved. Then we connect the word London in the tag cloud to a web site. Therefore we choose the menu entry *Connect visualization with websites* to enter the Mapping Editor. We choose the visualization and mark the word London with a mouse click. We now can enter a title and an URL of a website and save the connection. The visualizations now have multiple connections to other resources of which some were created automatically. The tag cloud contains links to Wikipedia for every word that represent a Wikipedia article, the location London in the map links to related web sites. The visualizations are linked to each other and additionally we have created a link from London to a related web site manually. Within minutes we have what could serve as a info graphic for tourism etc.

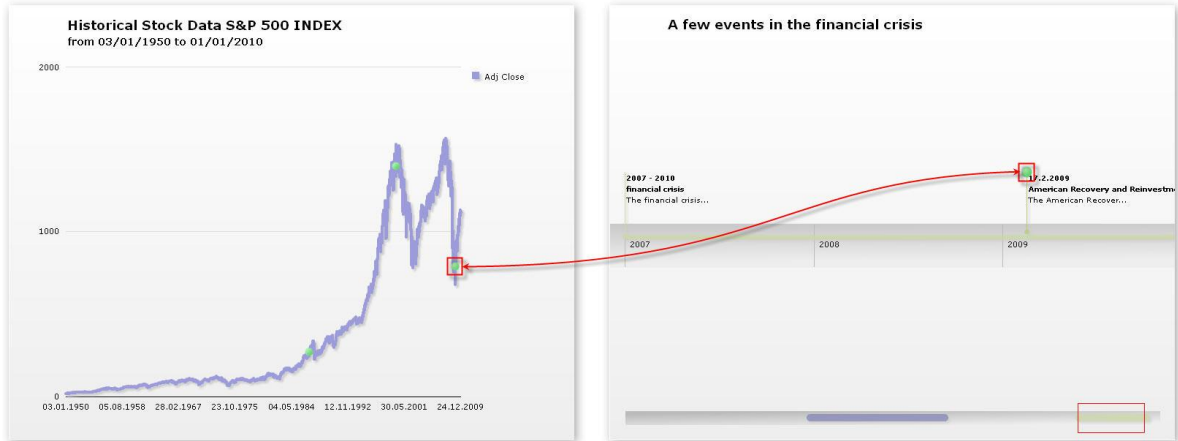

Figure 6: The left visualization presents historical stock prices of S&P 500 index in a line chart. The right visualization shows events on the financial crisis on a timeline. The user can browse from a minima of the line chart directly to the event on a timeline by clicking the green linking button.

## 5 EVALUATION

We have carried out a user study to approve that users can create visualizations simple and fast and can connect them to other visualizations or websites with the help of the Mapping Editor. We asked for a detailed assessment of each task, asked questions to approve the user has understood the concept and asked for scenarios.

### 5.1 Method

We gathered the participants of our user study by using the Mechanical Turk, an online workers marketplace offered by amazon.com. For 4US$ each, 100 participants were asked to complete 6 tasks on vizgr.com and then fill out an online survey about it.

We cleaned out all surveys that did not contain the basic information on demographics. Three of the surveys were only half-filled. As the corresponding accounts did not suggest technical difficulties (all tasks were fulfilled), we removed the unanswered question from the evaluation.

### 5.2 Demographics

The demographics seem quite typical for the internet population and surprisingly unbiased towards the archetypical heavy internet users (although we do have them). For the evaluation, we asked four demographical questions: Gender, Age, Education level and average internet usage.

Out of the valid surveys we received 53.6% were male and 46.4% female. The most common age group was 18 to 29 with 59.1%, followed by 32.7% 30 to 39. All participants have a high school degree; most (52.7%) even have a college degree. The average time spent on the internet was given at 29.4 hours per week, but varying from as little as 4 hours to an unrealistic 105 hours.

Although we did not explicitly ask for country of birth, the IP addresses used suggest that the vast majority (over 90%) of the users were logging in from the United States, with small minorities from a variety of rich first-world countries (Germany, Singapore, …).

### 5.3 Questions and Tasks

The participants were asked to complete four basic tasks on the Vizgr website. For each of the steps:

1. creating a tag cloud of the Wikipedia article on *London School of Economics*,
2. creating a map with Wikipedia data from *London*,
3. connecting both visualizations,
4. connecting the word *London* in the tag cloud to a web site,

we asked whether the participants had succeeded in the task, how long it took them, how difficult it was and how they would judge the usability of this in their life and in general.

The success rate was (in order): 97.3%, 94.5%, 87.7% and 87.6%. Failure was consecutive, all three that failed at the first step, also failed at the other steps, e.g. one person did not find the save button and was thus not able to finish any of the visualizations. When asked about obstacles, many participants found the first step "easy" or "simple", even though some of them never knew what a tag cloud was, before they entered the survey. There were some complaints about the save functionality or other minor user interface issues. 74.6% managed to finish the task in less than five minutes and 82.8% found it normal to very easy, compared to only 17.2%, who found it difficult or very difficult.

In the second step, creating a map, the three new failures were based on a misunderstanding. They had simply found and then used the prepared map for them, without even trying to create a new one. Only one of them also failed at the next steps. The commentary on this step was very mixed; some found it easier than the last step, others exactly the other way around. 75.5% finished in less than five minutes, 23.7% found it difficult or very difficult. This is somewhat comparable to the last step.

Connecting the visualizations seemed to be more difficult, with only 87.7% success rate. Besides the inherited failures from the tasks before, many participants were unsure on their actions, because they had no clear picture on what a successful link would look like, both visually and conceptually. This made some of the participants to give up on the task, but it also is the predominant theme in the commentary of the more successful participants. 81.1% of the participants managed to finish in less than five minutes, 21.7% found it difficult or very difficult. When we asked the participants, what the effect of the connection was, 69.8% of all participants explained it visually and/or conceptually. 11.9% either answered very generally ("it combines") or off-topic ("user friendly"); the rest (18.8%) either did not finish the task or did not understand the idea.

Although the success rate for linking the visualizations and linking a visualization to a web site are very similar, the participants failing were not identical to each other. No one seemed to be confused about the purpose of linking to a web page. The failures and negative comments were mostly due to technical or GUI problems. A common comment at that point was that fulfilling the steps before was helpful in understanding and executing this step as well. 83.8% finished in less than five minutes, 20.0% found it difficult or very difficult.

## 5.4 Conclusions

The great majority of the users had no or only small difficulties when working on the tasks. The difficulty of the different tasks seems to be stable when looking at time consumption and perceived difficulty. The failure rate seems to be much higher when being confronted with a new concept, be it tag clouds or connecting visualizations. The last step of connecting the visualization to a web site was technically the most difficult, counting e.g. the number of distinct mouse clicks. Yet, its conceptual simplicity ranked it near the technically much simpler task of connecting the visualizations.

A problem here seems to be the metaphor for the linking. About 10% of the users had not understood what the green dot or the red dot does in the visualization, although they had created the link themselves. This is a rather high rate, when compared to better known symbols, such as underlined text in a web site. But since the connection of visualization does not have such a standardized metaphor, we are quite satisfied with this number, although it does offer room for improvement.

The effect of the education level on the rate of failure or perceived difficulty is slight at best. While higher educated users were taking less time on the tasks and generally rated them easier, they also failed more often to complete the task at all. Also, unlike we conjectured, a general aptitude with the internet also did not help with completing the tasks. In fact, one of the three participants that were not able to complete any of the tasks was also number three when it came to internet usage, with 68 hours per week.

## 5.5 Scenarios

In last two steps about connecting visualizations, we asked the participants to give us scenarios, how they themselves or the general public could be using the tool for their benefit. 58.8% thought they could use the tool in their daily life, 35.6% could give an example, different to the one we were providing them and 10% could even give more than one example. 82% could see a use for the general public and 47.2% could give one or more examples of a use case, often different from the one they had given for the last question.

The scenarios ranged from general organization of thought over social linking, e-learning, tourism, marketing to the organization of knitting pattern and crime solving. Many pointed out that they found the layout and presentation appealing, so they would like to use it instead of traditional methods of presenting information. Scenarios similar to the one given in the example were most often. Most were connected to tourism, planning vacation or giving locality information to a friend or a guest. The second largest cluster of ideas was connected to e-learning, researching and presenting topics.

## 5.6 Feedback

We had a lot of general positive feedback at the end of the survey, several commented that they would like to use the tool in the future. Others were very eager to point out small bugs, spelling mistakes or the unfavourable colour scheme and promised to use the tool as soon as this was fixed. We received no general bad comments, just that two participants could see no benefit in Vizgr at all.

# 6 FUTURE WORK

As became apparent during the survey, not all of the participants immediately understood that the green dot was supposed to be the connection point between the two visualizations or to other resources. This is something we need to be working on; trying to find a more intuitive metaphor. We are planning to start another survey on this in the near future.

Apart from this and other minor GUI changes, we need to expand the tool into broader use cases. Different types of media should be connectable, not just data visualizations, but also images and annotated text e.g. from web sites.

The integration of semantic data sources like DBpedia, whose availability increases massively since the emerging Linked Open Data movement, has great benefits for the creation and linking of visualizations. Data like text, locations or events are imported user-friendly and fast. Linkings to other resources are already integrated and can be made visible on a visual level. We plan to integrate more semantic databases like freebase and want to offer a finer selection of resources.